\documentclass[%
 aip,
 jmp,%
 amsmath,amssymb,
reprint,%
]{revtex4-1}

\usepackage{graphicx}
\usepackage{epstopdf}
\usepackage{dcolumn}
\usepackage{bm}

\newtheorem{Guideline}{Guideline}

\begin{document}


\title{Unsupervised machine learning for detection of phase transitions in off-lattice systems I. Foundations}

\author{R. B. Jadrich} 
\author{B. A. Lindquist} 
\affiliation{McKetta Department of Chemical Engineering, University of Texas at Austin, Austin, Texas 78712, USA}
\author{T. M. Truskett} 
 \email{truskett@che.utexas.edu}
\affiliation{McKetta Department of Chemical Engineering, University of Texas at Austin, Austin, Texas 78712, USA}
\affiliation{Department of Physics, University of Texas at Austin, Austin, Texas 78712, USA}

\date{\today}

\begin{abstract}
We demonstrate the utility of an unsupervised machine learning tool for the detection of phase transitions in off-lattice systems. We focus on the application of principal component analysis (PCA) to detect the freezing transitions of two-dimensional hard-disk and three-dimensional hard-sphere systems as well as liquid-gas phase separation in a patchy colloid model. As we demonstrate, PCA autonomously discovers order-parameter-like quantities that report on phase transitions, mitigating the need for {\em a priori} construction or identification of a suitable order parameter--thus streamlining the routine analysis of phase behavior. In a companion paper, we further develop the method established here to explore the detection of phase transitions in various model systems controlled by compositional demixing, liquid crystalline ordering, and non-equilibrium active forces.

\end{abstract}

\maketitle

\section{Introduction} 

Phase transitions, diverse in character and ubiquitous in physical and biological systems, result from the correlated response of a near-infinite number of interacting microscopic degrees of freedom to a change in one or more macroscopic variables.~\cite{stat_mech_book_pathria,stat_mech_book_sethna,stat_mech_book_davis,noneq_pts_review,noneq_pts_book_v1,noneq_pts_book_v2} Here, we focus on off-lattice models of particle-based systems, i.e., atomic/molecular materials and their colloidal analogs, which are known to exhibit a rich variety of equilibrium phase transitions including condensation, freezing, and compositional demixing, as well as formation of microphases including cluster fluids.~\cite{simple_liquids_w_soft_matter} Phase transitions that occur due to nonequilibrium driving forces (e.g., oscillatory shear~\cite{oscill_shear_pts_1,oscill_shear_pts_2,oscill_shear_pts_3, oscill_shear_pts_4, driven_pts_ex}, time dependent magnetic/electric fields~\cite{oscill_mag_assembly, oscill_elec_mag_assembly}, or particle self-propulsion\cite{active_matter_pts,active_particles_review,active_pts_ex}) are also possible in these materials. Given the diversity of microscopic degrees of freedom and macroscopic outcomes exhibited by such systems, new types of phase transitions and corresponding states of matter can be challenging to detect or predict from first principles.

Typically, phase transitions are described via an abrupt change in a coarse-grained representation of the system--i.e., an order parameter (OP)--that is sensitive to the specific character of the transformation and can provide insights into its origin.~\cite{stat_mech_book_pathria,stat_mech_book_sethna,stat_mech_book_davis,noneq_pts_review,noneq_pts_book_v1,noneq_pts_book_v2} Given the sheer number and variety of known phase transitions in such systems, it is perhaps unsurprising that there is no universal choice for the OP. Even for equilibrium transitions with a well-developed statistical mechanical foundation and established OPs, detection and systematic characterization of phase transitions can be computationally inconvenient, requiring simulations in specific ensembles.~\cite{microphase_simulation,poly_hs_simulation,patchy_mix_sim,transition_matrix} As a result, simple and general strategies for identifying known phase transitions and discovering new ones would be welcome. 

Toward this end, one promising approach is the use of dimensionality reduction techniques from machine learning to extract OP-like descriptors for phase transitions from configurational data.~\cite{br_and_ml_book,stat_learning_book,machine_learning_in_physics_1,machine_learning_in_physics_2} For example, principal component analysis (PCA) has recently been used to detect phase transitions in the two-dimensional ferromagnetic Ising, XY and other related spin models.~\cite{spin_ml_1,spin_ml_2,spin_ml_3,spin_ml_4,spin_ml_5,spin_ml_6} Beyond PCA, research into nonlinear machine learning strategies has led to the development of 1) a so-called ``confusion''-based scheme which successfully detected transitions in the Kitaev Chain and Ising models, as well as in a disordered quantum spin chain,~\cite{spin_confusion_ml} and 2) neural networks that identified emergence of subtle many-body localized phases as well as those arising in the square-ice model and in Ising lattice gauge theory.~\cite{spin_neural_classification}

Here, we build upon this prior work and explore the extension of PCA to detect and characterize phase transitions for off-lattice models of particle-based materials. Specifically, we perform PCA on features derived from the particle coordinates, a strategy that is straightforward to implement in simulations but could also be applied in experiments that track particles (e.g., confocal microscopy of colloidal dispersions). In this first paper of a two-part series, we establish general guidelines and practices for using PCA to detect phase transitions in such systems. In the second part, we test the generality of this strategy for a variety of particle-based systems that exhibit equilibrium and non-equilibrium phase transitions.

The balance of this manuscript is organized as follows. In Section~\ref{sec:methods}, we describe the simulation protocols utilized in this work, provide a brief review of PCA, and identify and resolve complications that arise from a na\"{i}ve construction of features from particle coordinate data. In Section~\ref{sec:results}, we present the results of PCA analysis of the features for some canonical phase transitions: the freezing transition in monodisperse hard disks and hard spheres and spinodal decomposition in a three-dimensional binary mixture with complementary attractions. We present conclusions for this study in Section~\ref{sec:conclusions}.

\section{Computational Methods}
\label{sec:methods}
\subsection{Simulation and analysis}
Hard-disk and hard-sphere simulations were performed using the hard-particle Monte Carlo integrator in the HOOMD-blue simulation suite,~\cite{hoomd_1,hoomd_2,hoomd_3} where $N=4096$ particles were simulated in periodically replicated square and cubic simulation boxes of side-length $L$, respectively. Random particle translations were restricted to 10-30\% of the particle diameter ($d$) along any Cartesian direction. These equilibrium hard-particle model systems are athermal and as such are uniquely described by an area or volume fraction ($\eta$); $\eta\equiv (N/L^{2}) \pi d^{2}/4$ for disks and $\eta\equiv (N/L^{3}) \pi d^{3}/6$ for spheres. For the PCA, we generated equilibrium samples over a range of $\eta$ spanning below and above the freezing transition via a successive compression, equilibration, and sampling scheme. For hard disks we incremented $\eta$ from 0.55-0.80 in increments of 0.005, performing $1\times10^{7}$ HOOMD ``timesteps'' (where a timestep is approximately equal to four sweeps of MC moves over all particles) at each state point until $\eta=0.695$ and $5\times10^{7}$ timesteps per state point thereafter. Between $10^3$ and $10^4$ evenly spaced configuration snapshots were stored at each state point to provide sufficient data for the PCA analysis. For hard spheres, samples were generated for packing fractions between 0.40-0.70 in increments of 0.005. Below $\eta=0.53$, $10^6$, time steps were used, and 200 evenly spaced snapshots were stored. Above $eta=0.53$, 10 separate trajectories were randomly initialized and run for $5\times10^5$ time steps while only storing 20 evenly spaced configurations per trajectory.

For the case of spinodal decomposition, we revisit simulation data acquired in Ref.~\citenum{linkergel}, where we studied binary mixtures of spherical particles decorated with six patches and two patches, respectively. While the spherical cores exhibited excluded volume interactions, patches on unlike particle types were mutually attractive. More details on the model and simulation protocol for the patchy particle mixture are described in Ref.~\citenum{linkergel}. In this work, we re-analyze the simulation data for the ratio of the number of two-patch particles to six-patch particles, $\gamma$, equal to 1.3. We use PCA to detect liquid-gas phase separation and compare the results to the ($\eta$, $T$) phase diagram reported in Ref.~\citenum{linkergel} on the basis of $k=0$ structure factor calculations.

In all cases, PCA was carried out using the Incremental PCA (IPCA) module in Scikit-Learn.~\cite{sklearn} Importantly, IPCA allows for training using only small portions of the full data set at once; this is useful given the large size of the simulation trajectories aggregated across state points.

All visualizations were rendered using the Open Visualization Tool (or OVITO).~\cite{ovito} Hard sphere renderings utilized the built-in Polyhedral Template Matching (or PTM) capabilities of OVITO for crystal structure identification.~\cite{ptm}

\subsection{Principal component analysis: Review}
\label{subsec:pca_review}
Principal component analysis (PCA) is a widely used unsupervised machine learning method that systematically discovers a lower dimensional representation of high dimensional data.~\cite{br_and_ml_book,stat_learning_book,pca_guide} Here, we briefly review the general concepts behind PCA. Interested readers can find more detailed descriptions in the literature. 

PCA, like all machine learning models, requires a numerical description for each of the $M$ observations that describe the system. A given observation is represented by a so-called ``feature'' vector 
\begin{equation} \label{eqn:feature_vector}
\boldsymbol{f} \equiv \begin{bmatrix}
    f_{1},   & \dots,  & f_{m}     
\end{bmatrix}^{T}
\end{equation} 
where each of the features ($f_{i}$) is a scalar measurable, $T$ denotes a transpose, and $m$ is the dimensionality of the feature representation. For context, observations in our work are derived from snapshots of particle configurations. The feature vector could describe the entire configuration or a subset of the particles. Without loss of generality, we assume that the elements of the feature vectors are centered (i.e., their mean value is zero). In most applications $m \gg 1$. To conveniently denote a data set of observations (implicitly of size $M$), we adopt the notation $\mathcal{D}$. 

If multiple features are highly correlated across the $M$ observations, then the dataset necessarily contains redundant information. Dimensionality reduction strategies provide a mechanism for ``concentrating'' this redundant information into fewer dimensions. Towards this end, PCA determines a linear transformation that yields a decorrelated feature representation of the data. The transformation provides a new set of $m$ directional unit vectors $\boldsymbol{w}_{i}$ (of length $m$) that define a new coordinate system and a corresponding $m\times m$ transformation matrix
\begin{equation} \label{eqn:loading_matrix}
\boldsymbol{W} \equiv \begin{bmatrix}
    \boldsymbol{w}_{1},  & \dots,  & \boldsymbol{w}_{m}   
\end{bmatrix}^{T}
\end{equation}
The projection of $\boldsymbol{f}$ along any of these new directions is simply $p_{i}=\boldsymbol{w}_{i}^{T} \boldsymbol{f}$. This is more succinctly stated in terms of the transformation matrix and the vector of $m$ projections, $\boldsymbol{p}\equiv \begin{bmatrix}p_{1}, & p_{2}, & \dots, & p_{m}\end{bmatrix}$, via
\begin{equation} \label{eqn:pca}
\boldsymbol{p} = \boldsymbol{W} \boldsymbol{f}
\end{equation}
For linear decorrelation, $\boldsymbol{W}$ must be of a form that diagonalizes the covariance matrix of $\boldsymbol{p}$
\begin{equation} \label{eqn:covariance}
\begin{split}
& \langle \boldsymbol{p}\boldsymbol{p}^{T} \rangle_{\mathcal{D}} = \boldsymbol{W} \langle \boldsymbol{f}\boldsymbol{f}^{T} \rangle_{\mathcal{D}} \boldsymbol{W}^{T}\\
& \langle p_{i}p_{j} \rangle_{\mathcal{D}} = 0, \ \ i\neq j
\end{split}
\end{equation} 
where $\langle \cdots \rangle_{\mathcal{}D}$ denotes an average over the data. 

Given the above prescription, $\boldsymbol{W}$ is still underdetermined. Uniqueness is realized within PCA by requiring that $\boldsymbol{W}$ be an orthogonal matrix ($\boldsymbol{W}^{T}=\boldsymbol{W}^{-1}$, where the right hand side notation indicates an inverse matrix). Under this constraint, the orthogonal set of unit vectors contained in $\boldsymbol{W}$ is equivalent to the sequential construction of $\boldsymbol{w}_{i}$ such that the variance projected along them ($v_{p_{i}}$) is maximized while maintaining orthogonality with all previous vectors $\boldsymbol{w}_{i}\cdot\boldsymbol{w}_{j}=0$ for all $j<i$. The $v_{p_{i}}$ are equal to the eigenvalues of the diagonalized covariance matrix $\langle \boldsymbol{p}\boldsymbol{p}^{T} \rangle_{\mathcal{D}}$.

The $\boldsymbol{w}_{i}$ constructed in the aforementioned manner are denoted PCs and the corresponding $p_{i}$ are called PC scores. Each PC contains more information than the next, as quantified by the variance (i.e., $v_{p_{1}} \geq v_{p_{2}} \geq \dots \geq v_{p_{m}}$). The efficiency with which information is sequestered towards the earlier PCs can be quantified by computing the relative explained variance $\lambda_{i}$ of each component 
\begin{equation} \label{eqn:explained_variance}
\lambda_{i} = \dfrac{v_{p_{i}}}{\sum_{j=1}^{m}v_{p_{j}}}
\end{equation}
The data is generally considered amenable to dimensionality reduction by this approach when there is a small dominant set of PCs with large relative explained variance.

\subsection{Principal component analysis: Order parameters for particle based systems}

The primary challenge associated with application of PCA to particle-based systems is feature construction. Uninformative features will yield poor results regardless of the level of sophistication of the machine learning model. On the other hand, informative features can enable even the most restrictive methods to detect useful correlations. Thus, application of machine learning to physics problems has distinct advantages: both intuition about the system as well as fundamental physical requirements can be incorporated to construct better features. 

In prior work that applied PCA to spin-based systems, raw configuration data has often been used directly to construct the feature vectors.~\cite{spin_ml_1,spin_ml_2,spin_ml_3} Indeed, raw particle coordinates could na\"{i}vely be employed as features in this work as well. However, particle-based systems are described by continuous variables (instead of discrete states as for spins), necessarily leading to a sparser data set given the same number of features. Physically motivated rotational and translation symmetry constraints can be employed during feature construction to greatly reduce the sparsity of the data.  

One of the simplest sets of features that encode invariance to translation and rotation is the array of interparticle distances about a random tagged particle; however, choices regarding sorting of the distances can play an important role. We observe that sorting on the basis of nearest neighbor (NN) distance is necessary for PCA to robustly detect a phase transition. Sorting assigns a well-founded physical interpretation to each feature that is probed across separate observations. We also choose to nondimensionalize the distances by the mean interparticle separation ($l\equiv\rho^{-1/D}$, where $\rho$ is the number density and $D$ is the spatial dimension). For this paper, our features are thus
\begin{equation} \label{eqn:nearest_neighbors}
\begin{split}
&\boldsymbol{\delta r} \equiv \begin{bmatrix}
    \delta r_{1},      & \delta r_{2}, & \dots,  & \delta r_{m}     
\end{bmatrix}^{T}, \\
&\delta r_{i} \equiv r_{i} - \langle r_{i} \rangle_{\mathcal{D}}, \\
&r_{1} \leq r_{2} \leq \dots \leq r_{m}
\end{split}
\end{equation} 
where $r_{i}$ is the (implicitly non-dimensionalized) $i^{\text{th}}$ nearest neighbor (NN) distance about a probe particle, sorted in ascending order. We refer to this encoding as the NN representation. Features like $\boldsymbol{\delta r}$ that encode some basic physical intuition are more generally referred to as intuited features, and are denoted by $\boldsymbol{f}_{\text{I}}$ 

As described in the previous subsection, PCA concentrates the variance in the data into a lower dimensional space, regardless of the origin of that variance. That is, PCA does not ``know'' to produce an OP. Therefore, PCA will be best suited to detect a phase transition when the bulk of the variance is due to the phase transition itself and not other factors. Therefore, it is instructive to consider the behavior of the feature vector in the absence of any effects connected to tuning any of the thermodynamic variables. For this work, a useful reference state is the low density ideal gas limit,~\footnote{For phase transitions where the disordered phase does not possess a true ideal gas limit, a different reference state may be warranted.} which has no intrinsic positional correlations. Since raw ideal gas positional data does not contain redundancy, in principle, no dimensionality reduction should be possible on such data (i.e., the explained variance of each of the $i$ dimensions, $\lambda_i$, should be flat).

\begin{figure}
  \includegraphics[width=3.37in,keepaspectratio]{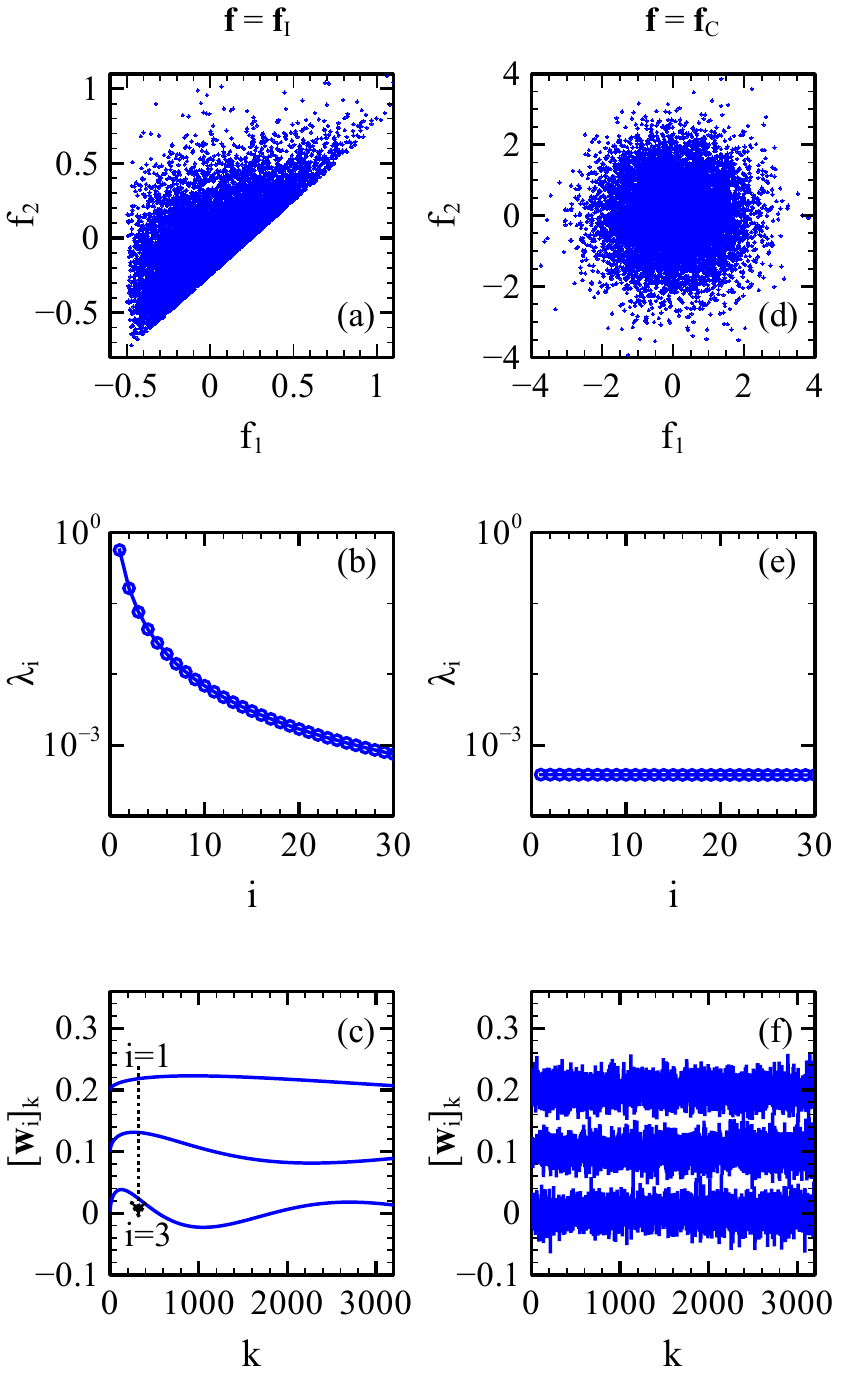}
  \caption{Application of PCA to NN encoding of ideal gas data using the first 3200 NNs. (a-c) 1000 samples in $f_{1}-f_{2}$ space, relative explained variance, and PC weights for intuited NN features, respectively. (d-f) The same as for a-c but using corrected NN features, where the individual features have been shifted for clarity.}
  \label{fgr:intuited_features_ig}
\end{figure}

However, the above sorting scheme, where NN distances are placed in the feature vector in ascending order, artificially imbues the features with trivial correlations: $f_{2} \ge f_{1}$, for instance, as shown in Fig.~\ref{fgr:intuited_features_ig}a. As a result, PCA ``learns'' about these correlations and discovers a transformation for which the values for $\lambda_i$ vary strongly (Fig.~\ref{fgr:intuited_features_ig}b)--an indication of meaningful dimensionality reduction on the reference ideal gas data set. As PCA is explicitly probing strong co-variation among the features to identify a more informative coordinate system, it is reasonable to expect any underlying phase transition to be obscured by such trivial correlations in the features. Indeed, preliminary PCA calculations were complicated by such effects when the features $\boldsymbol{f}_{\text{I}}$ as defined by Eq.~\ref{eqn:nearest_neighbors} were employed. Physically, the trivial NN correlations correspond to NN packing modes of increasing frequency, as shown by the first three PC weights in Fig.~\ref{fgr:intuited_features_ig}c.

To prevent trivial correlations from obscuring--and possibly mixing with--those that arise from a phase transition, as well as to recover the expected ideal gas behavior upon application of PCA, we transform an arbitrary $\boldsymbol{f}_{\text{I}}$ into corrected features ($\boldsymbol{f}_{\text{C}}$) that are truly uninformative in the reference state. In this work, the transformation $\boldsymbol{f}_{\text{I}}\rightarrow\boldsymbol{f}_{\text{C}}$ is constructed via a linear ``whitening'' transformation~\cite{whitening} derived from reference system data ($\mathcal{D}_{0}$). A whitening transformation performs a change of basis to approximately convert correlated data into white noise, for which the co-variance matrix must be the identity matrix. Here we provide one possible prescription that we adopt to whiten the data: a PCA transformation to decorrelate the data followed by normalization of the resulting $\boldsymbol{w}_{i}$ by the square root of the variance (i.e., the standard deviation) along them, as follows
\begin{equation} \label{eqn:pca_whitening}
\boldsymbol{f}_{\text{C}} = \boldsymbol{W}_{0} \boldsymbol{f}_{\text{I}}
\end{equation}
where the orthogonal matrix $\boldsymbol{W}_{0}$ yields the unit or identity matrix ($\boldsymbol{I}$) in the reference state
\begin{equation} \label{eqn:whitening_covariance}
\boldsymbol{I} = \langle \boldsymbol{f}_{\text{C}}\boldsymbol{f}_{\text{C}}^{T} \rangle_{\mathcal{D}_{0}} = \boldsymbol{W_{0}} \langle \boldsymbol{f}_{\text{I}}\boldsymbol{f}_{\text{I}}^{T} \rangle_{\mathcal{D}_{0}} \boldsymbol{W_{0}}^{T}
\end{equation} 
The $\boldsymbol{f}_{\text{C}}$ can be qualitatively regarded as a linear combination of whitened versions of the ``packing mode'' PCs found in Fig~\ref{fgr:intuited_features_ig}c. 

In the ideal gas limit, the $\boldsymbol{f}_{\text{C}}$ approximate Gaussian white noise (Fig.~\ref{fgr:intuited_features_ig}d) and yield an uninformative explained variance spectrum (Fig.~\ref{fgr:intuited_features_ig}e). In addition, the PC weights, shown in Fig.~\ref{fgr:intuited_features_ig}f, possess no discernible structuring. Thus, when using $\boldsymbol{f}_{\text{C}}$ to encode data outside of the reference, anything discovered by PCA is effectively in ``excess'' of the reference, by analogy to thermodynamics. 

PCA is used here to automatically construct an OP-like quantity for the phase transition exhibited in the simulation data. In particular, we explore the utility of $p_{1}$, the score associated with the most informative principal component, as the OP. As the $p_{i}$ are defined at the feature vector level, ensemble-averaged analogs $P_{i} \equiv \langle p_{i} \rangle_{\mathcal{S}}$ are defined so that the average is implicitly limited to data at a single state point $\mathcal{S}$. 

Using the component weights, it is sometimes possible to ascribe a qualitative physical interpretation for the OP as the weights directly reflect the relevance of the features. To interpret the components in terms of the physically intuitive features, $\boldsymbol{f}_{\text{I}}$, as opposed to $\boldsymbol{f}_{\text{C}}$, we require the weights governing the full transformation $\boldsymbol{f}_{\text{I}} \rightarrow \boldsymbol{p}$. Substituting Eqn.~\ref{eqn:pca_whitening} into Eqn.~\ref{eqn:pca} with $\boldsymbol{f}=\boldsymbol{f}_{\text{C}}$ we arrive at the full $m \times m$ transformation matrix
\begin{equation} \label{eqn:nested_matrix}
\boldsymbol{Q}  \equiv \boldsymbol{W} \boldsymbol{W}_{0} \equiv\begin{bmatrix}
    \boldsymbol{q}_{1}, & \boldsymbol{q}_{2}, & \dots,  & \boldsymbol{q}_{m}
\end{bmatrix}^{T}
\end{equation}
where the $\boldsymbol{q}_{i}$ are column vectors of length $m$ containing the effective ``nested'' weights mapping each scalar feature in $\boldsymbol{f}_{\text{I}}$ to $p_{i}$.

Based on the observations of this section, we propose two guidelines for employing PCA as a tool to discover phase transitions in off-lattice models of particle-based systems:
\begin{Guideline}
Choose a representation in which translations and rotations of a given configuration do not alter the corresponding feature description.
\label{guidline:1}
\end{Guideline}
\begin{Guideline}
Ensure the chosen feature representation of the data corresponds to uncorrelated features of equal variance in the disordered reference state.
\label{guidline:2}
\end{Guideline}
Specific to the latter guideline, we found linear whitening to be sufficient for PCA, as it is a linear learner. More sophisticated learners may require nonlinear whitening.

\section{Results}
\label{sec:results}

\subsection{Freezing}
As a first test of PCA for detecting phase transitions in off-lattice particle-based systems, we examine the well-known ordering transitions exhibited by rigid radially symmetric particles in two and three dimensions (i.e., hard-disk and hard-sphere models, respectively) upon densification of the disordered fluid.

\begin{figure}
  \includegraphics[width=3.37in,keepaspectratio]{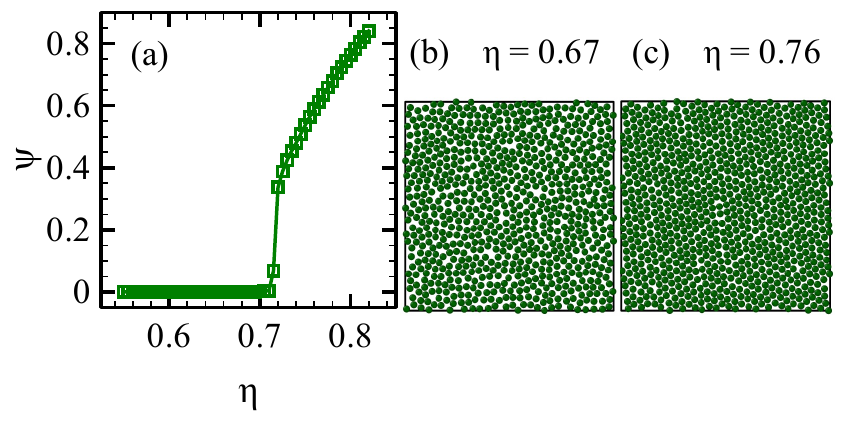}
  \caption{(a) Behavior of the classical positional OP for freezing in hard disks, $\psi$, versus packing fraction, $\eta$, with example (b) fluid and (c) triangular solid configurations bracketing the transition.}
  \label{fgr:hd_classical_op}
\end{figure}

In the macroscopic limit, the hard-disk fluid does not directly freeze into a triangular crystalline solid upon densification; there is an intermediate hexatic phase that emerges in a narrow window of density and requires extremely large-scale simulations to resolve.~\cite{hard_disks_1,hard_disks_2} For simulations of moderately-sized systems such as those considered here, the transition manifests as weakly first-order between the fluid and the triangular solid. Simulated assembly of the solid upon densification of the fluid is facile and the resulting structure is notably reproducible, in contrast to that of analogous three-dimensional hard-sphere systems (as discussed below). The positional OP traditionally employed to detect two-dimensional freezing of hard disks, $\Psi$, is defined by the following equation,
\begin{equation} \label{eqn:classical_pop}
\Psi \equiv \bigg| \dfrac{1}{N}\sum_{i=1}^{N}e^{i\boldsymbol{k}\cdot\boldsymbol{r}_{i}}   \bigg|^{2},
\end{equation} 
where $\boldsymbol{r}_{i}$ is the vector position of particle $i$, $N$ is the number of particles, and $\boldsymbol{k}$ is the reciprocal lattice vector for the relevant crystal.~\cite{hard_disks_op_1,hard_disks_op_2} The magnitude of the reciprocal lattice vector for a hard-disk triangular lattice depends on density ($\rho=4\eta/\pi d^{2}$) as: $|\boldsymbol{k}|\approx 2\pi / \sqrt{\sqrt{3}/2\rho}$. Here, we choose the vector orientation so as to maximize $\Psi$ at the densest state point ($\eta=0.82$), the results of which are shown in Fig.~\ref{fgr:hd_classical_op} along with two snapshots of particle configurations at $\eta$ below and above the freezing transition, respectively. $\Psi$ rapidly increases from zero to finite values in the narrow density window $\eta=0.71-0.72$ with continued growth as the ordered phase is densified. From a physical standpoint, $\Psi$ probes the emergence of long-ranged oscillatory packing correlations with a wavelength set by mean first coordination shell distance in the triangular lattice.

\begin{figure}
  \includegraphics[width=3.37in,keepaspectratio]{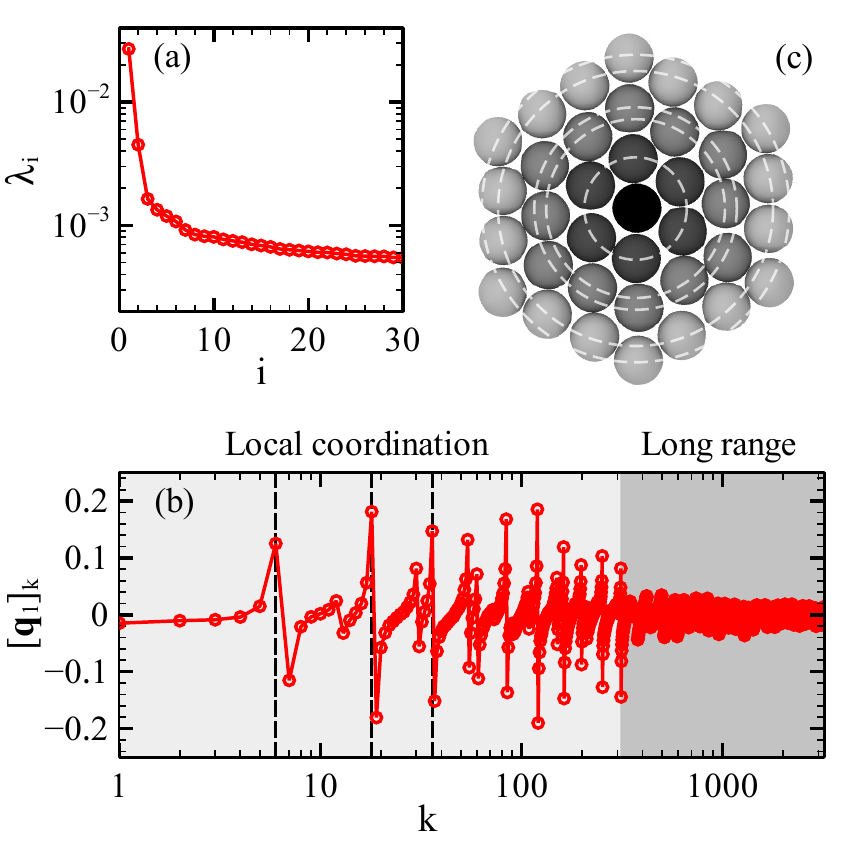}
  \caption{Application of PCA to the hard-disk model data using corrected NN features with 3200 NNs in total. (a) Explained relative variance. (b) Nested weights as a function of the NN index for the first principal component. Vertical lines indicate the completion of the first, second, and third coordination shells respectively. (c) Schematic of the first three completed coordination shells in the triangular lattice of hard disks.}
  \label{fgr:corrected_features_hd}
\end{figure}

\begin{figure}
  \includegraphics[width=3.37in,keepaspectratio]{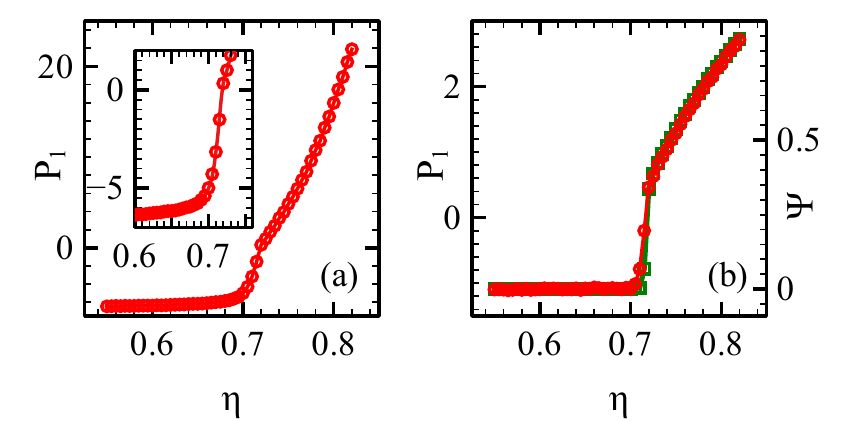}
  \caption{PCA-deduced OP for hard disks using the corrected NN encoding (a) with all of the first 3200 NNs and (b) only including every $100^{\text{th}}$ NN in the feature vector and overlaid with the classical OP ($\Psi$) from Fig.~\ref{fgr:hd_classical_op} (green squares)}
  \label{fgr:hd_pca_op}
\end{figure}

Application of PCA to the corrected NN representation of the same data used for the classical OP calculation yields a dominant PC ($p_{1}$), as assessed by $\lambda_{i}$ in Fig.~\ref{fgr:corrected_features_hd}a. As depicted in Fig.~\ref{fgr:corrected_features_hd}b, the effective weights associated with the first PC can be roughly separated into local and long-range contributions. The former encode strong coordination shell structuring, while the latter capture the long-range positional correlations inherent to the crystal. The interpretation of the local structure can be further clarified by examining the dominant positive and negative weight pairs observed in Fig.~\ref{fgr:corrected_features_hd}b. Specifically, these signatures emerge at length scales corresponding to successive coordination shells. For a triangular lattice, the $n^{\text{th}}$ coordination shell is comprised of $6n$ particles. Therefore, the first three shells correspond to 6, 18, and 36 nearest neighbors in total, as indicated by the vertical lines in Fig~\ref{fgr:corrected_features_hd}b--a corresponding visual interpretation is provided in Fig.~\ref{fgr:corrected_features_hd}c. Weak splitting of the second and third shells into two sub-populations is also captured by the more subtle intermediate jumps in the weights.

The PCA-derived OP shown in Fig.~\ref{fgr:hd_pca_op}a is behaviorally similar to the classical analog in Fig.~\ref{fgr:hd_classical_op}, with a relatively flat region at low densities, sharp growth at the freezing transition, and more muted, but continued growth, at higher densities. The preceding qualitatively trends are not particularly sensitive to the exact definition of the feature vector. For instance, we find that several modifications of the original feature vector--by only including 1) every tenth NN distance, 2) every $100^{\text{th}}$ NN distance, or 3) the long-ranged portion of the original feature vector--all yield an identical range in $\eta$ for the phase transition (data not shown). 

On a quantitative level, enhanced agreement between the PCA-deduced and traditional OPs is realizable by sub-sampling the feature construction as described above. While the classical OP only probes the growth of long-ranged oscillatory packing correlations characteristic of the triangular lattice, the PCA OP using the original feature vector captures both specific local structural details and generic long-ranged correlations as can clearly be seen in Fig.~\ref{fgr:corrected_features_hd}b. However, the above sub-sampling schemes remove the local coordination shell structuring from the PCA. For instance, by only including every $100^{\text{th}}$ NN distance in the feature vector, PCA discovers an OP that is remarkably similar to the classical analog, enabling a near perfect collapse of the two in Fig.~\ref{fgr:hd_pca_op}b. 

\begin{figure}
  \includegraphics[width=3.37in,keepaspectratio]{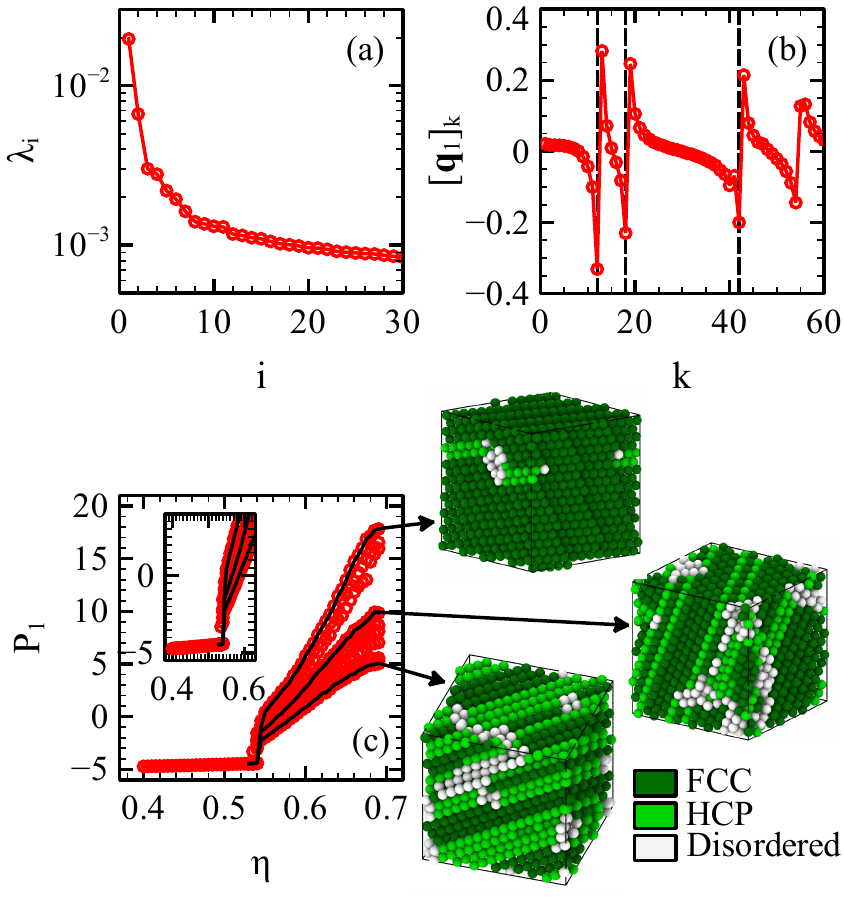}
  \caption{Application of PCA to hard spheres using the corrected NN encoding and the first 2100 NNs. (a) Explained relative variance for hard-sphere system. (b) Nested weights as a function of the NN features for the first principal component. (c) Branches of the PCA-deduced OP and corresponding snapshots of the solid phase at $\eta=0.68$; Polyhedral template matching~\cite{ptm} was used to identify crystal type.}
  \label{fgr:hs_pca_op}
\end{figure}

The PCA calculations for the hard-disk model can be trivially extended to detect the ordering transition in the three-dimensional hard-sphere model, for which there are various detection schemes in the literature.~\cite{steinhardt,cnn,ptm,vta,ml_id_1,ml_id_2} However, key practical differences regarding crystallization processes in these two systems are worth considering. First, once hard spheres crystallize in a simulation, there is an almost complete arrest of appreciable collective particle rearrangements. Defects or imperfect structures persist for long time scales in the assembled structure. Second, the ground state face-centered-cubic (FCC) lattice has a strong competitor, the hexagonal-close-packed (HCP) lattice, often leading to assembled structures that resemble a mosaic of FCC and HCP domains. Such domains can be identified via established structure identification methods.~\cite{cnn,ptm,vta,ml_id_1,ml_id_2} Annealing out any resultant mosaic pattern is practically impossible due to the aforementioned kinetic and thermodynamic issues. Robust PCA detection of the ordering transitions should reflect these practical differences.

To capture variations in the crystalline structures that naturally emerge upon densification of the hard-sphere fluid in simulations, our PCA analysis considers ten separate trajectories that are compressed through the freezing transition. The trained PCA model incorporates all of the data aggregated across each density and each trajectory--the outcome is a dominant PC (Fig.~\ref{fgr:hs_pca_op}a) with a weight vector (Fig.~\ref{fgr:hs_pca_op}b) that, like that for hard disks, captures the successive completion of coordination shells. Though not shown, long-range correlations also contribute to $P_{1}$. While PCA utilizes the aggregated trajectory data, computation of $P_{1}$ in Fig.~\ref{fgr:hs_pca_op}c is performed on an individual trajectory basis, yielding distinct OP branches for each. The overall behavior is similar to that found for hard disks: $P_{1}$ is relatively flat in the disordered phase and is followed by a marked increase near upon freezing with continued growth in the solid phase. Growth is maximal for the almost perfect FCC sample and markedly reduced for the alternating layered FCC+HCP structure. Mixing of two crystalline types necessarily detracts from the long range correlations, thus lowering the overall magnitude of $P_{1}$ in the solid phase. Relative to the known phase behavior of hard spheres, our transition appears to correspond to freezing from the overcompressed metastable fluid state, as the true freezing transition at $\eta=0.49$ is bypassed. Instead, the system freezes around the known melting point at $\eta=0.55$, where fluid-crystal coexistence ends in favor of a single phase crystal. 

\subsection{Liquid-gas Phase Separation}
\label{subsec:spinodal}

\begin{figure}
  \includegraphics{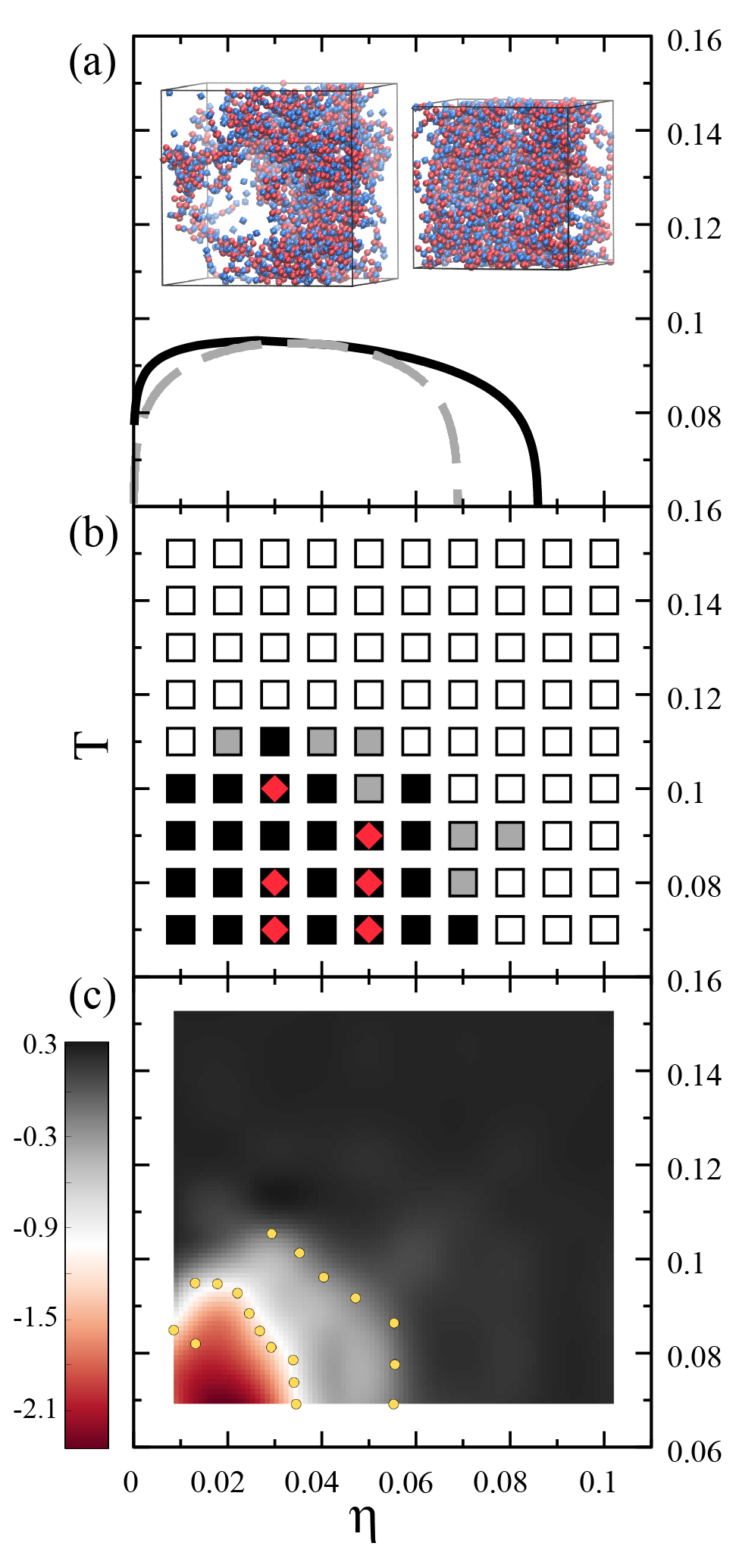}
  \caption{(a) Theoretical prediction for the binodal (solid line) and spinodal (dashed line) boundary for a patchy binary mixture described in Ref.~\citenum{linkergel}. (inset) Simulation snapshots from the phase-separated (left) and the homogeneous (right) regime, respectively. (b) Simulated phase diagram corresponding to panel (a), where solid black squares had $S(k=0)>10$, black squares with red diamonds had $S(k=0)<10$ but the variance metric indicated that they were phase-separated, gray squares had $5<S(k=0)<10$, and white squares had $S(k=0)<5$. (c) The OP derived from PCA of the simulation data used to generate panel (b). The yellow dots indicate the first two inflection points along lines beginning at the lower-left hand corner of the plot and at an angle $\theta$ relative to the x-axis. Panels (a) and (b) were adapted from Ref.~\citenum{linkergel} with permission.}
  \label{fgr:spinodal}
\end{figure}

To further test the utility of PCA for detecting phase transitions, we revisit our previously conducted study on the phase behavior of a binary mixture of patchy particles with complementary attractions in the canonical ensemble.~\cite{linkergel} Thermodynamic predictions of liquid-gas phase separation in such systems are often used as a surrogate for predicting gel formation, as the phase separation can be kinetically arrested to form a gel. Of particular interest is the binodal curve, which defines the boundary between conditions where a homogeneous fluid is the stable phase versus where a phase-separated (liquid and gas) state is thermodynamically preferred. Sometimes the easier-to-compute spinodal boundary, below which a single metastable fluid phase becomes unstable and spontaneously separates into a liquid and a gas, is determined instead. For colloids dispersed in a solvent, ``liquid'' refers to a particle-rich phase and ``gas'' to a particle-poor phase. However, for in silico models of colloids, solvent effects are often treated implicitly, and so the simulation configurations are more similar to a canonical liquid-gas delineation; see a phase-separated configuration (left) and a homogeneous configuration (right) in the inset of Fig.~\ref{fgr:spinodal}a.  

In Ref.~\citenum{linkergel}, we computed the binodal and spinodal boundaries for the above binary mixture as a function of packing fraction $\eta$ and temperature $T$ using Wertheim theory;~\cite{wertheim_1,wertheim_2,wertheim_3} see the curves in Fig.~\ref{fgr:spinodal}a. In order to validate the theoretical calculations, we performed a series of simulations, varying $\eta$ and $T$, over the region indicated by the theory. As described in the Introduction, many types of phase transitions can be appropriately modeled via complex simulations in specific ensembles.~\cite{microphase_simulation,poly_hs_simulation,patchy_mix_sim,transition_matrix} However, such simulations are generally computationally expensive and can be more difficult to implement in standard simulation packages. Therefore, in the case of spinodal decomposition, divergence of the structure factor at zero wavenumber $k=0$ (i.e., the compressibility) within a simulation in the canonical ensemble is often used to detect phase separation,~\cite{sk_spinodal,liquid_state_theory} which is the approach that we adopted in Ref.~\citenum{linkergel}. However, the low $k$ region is reflective of long-ranged behavior in real-space, rendering such calculations sensitive to finite-size effects. For instance, in the above study, we encountered cases where the compressibility was significantly underestimated as the correlation length grew too large relative to the simulation box, and so we needed to perform a secondary calculation quantifying the variance in density relative to a hard sphere reference to fully determine the phase boundary.~\cite{linkergel}  

Adapted from Ref.~\citenum{linkergel}, the results of the above simulation analysis for a single composition in the binary mixture are shown in Fig.~\ref{fgr:spinodal}b, where black squares represent simulations where $S(k=0)>10$, black squares with red diamonds indicate simulations for which $S(k=0)<10$ but the variance criteria indicated phase separation, gray squares represent $5<S(k=0)<10$, and white squares were determined to be homogeneous by all metrics. In total, we performed 90 simulations. As an alternative to the above analysis, we have applied PCA to the simulation data. Instead of analyzing the entire trajectory, we only included a single snapshot from each simulation, constructing a separate feature vector for each of the 2300 particles in the simulation. To accommodate this more modest sampling, we reduced the dimensionality of the feature vector by only including every tenth nearest neighbor after the first. The resulting OP from the PCA is shown as a heat map in Fig.~\ref{fgr:spinodal}c, where the data has been fit to a second-order bivariate spline. There appear to be three distinct regions that can roughly be identified by color--red, light gray, and black. The first two inflection points along lines emanating from the origin at an angle $\theta$ from the x-axis in increments of 10$^{\circ}$ are shown as yellow points (the scans for $\theta=80^{\circ}$, $90^{\circ}$ only have one inflection point). By comparing to simulation data above, we identify the red and gray regions as phase-separated and the black region as homogeneous. While the agreement is not perfect, the $S(k)>10$ criteria is approximate, and PCA may very well provide a more accurate classification of the data. Regardless, the qualitative comparison between theory and experiment is the same: the range of the phase-separated regime is slightly reduced in density but extends to slightly higher temperatures than the theory predicts. 

The presence of three distinct regions, instead of two, in the above analysis would somewhat complicate the use of PCA to identify the binodal if the conventional OP was not provided for comparison in Fig.~\ref{fgr:spinodal}b. However, the physical interpretation of the regions identified by PCA can be determined by more comprehensive calculations at just a few locations within the regions of interest, as opposed to screening the entire phase space in that fashion. Interestingly, we note that the two rough boundaries represented by the yellow circles in Fig.~\ref{fgr:spinodal}c bear qualitative resemblance to the binodal and spinodal curves shown in Fig.~\ref{fgr:spinodal}a. Since it is difficult to maintain equilibrium in the simulation once phase separation begins, there could be long-lived structural differences in simulated dense phases formed via decomposition (inside the spinodal) versus via nucleation (inside the binodal but outside the spinodal) that PCA is able to detect. 

Even if a handful of more intensive simulations are required, PCA represents a significant reduction in computational cost for the determination of the binodal. In addition to a more computationally efficient analysis (under one minute on a desktop computer for the PCA--several orders of magnitude shorter than those required to estimate the low-k behavior of the structure factor), the PCA calculations only required considering a single configuration snapshot (versus several thousand to compute the compressibility), drastically reducing the simulation time required to collect sufficient statistics.  

\section{Conclusions}
\label{sec:conclusions}
We have have laid the groundwork for--and demonstrated the application of--a general machine learning approach to automatically derive OPs. We showed that, in addition to translational and rotational invariance, it is important to ensure that the features used to encode the particle configurations are uninformative in the ideal gas limit. Without this, trivial correlations in the features can swamp out and/or mix with the contributions directly coming from any phase transition. We also demonstrated that features that do not respect this limiting behavior can be easily transformed into features that do via a whitening transformation derived from an equivalent ideal gas data-set. We anticipate that the best practices established in this work are generally extensible to more sophisticated machine learning models.

As a proof-of-principle study, we successfully applied PCA to detect the freezing transition in hard disks in two-dimensions and in hard spheres in three-dimensions. We also showed how PCA can be used to drastically reduce the computational expense of routine simulation analysis within the context of phase separation of a binary mixture of patchy particles possessing complementary attractions.


In the following companion article, we extend the framework developed here to other model systems that are known to possess other classes of phase transitions in order to evaluate the generality of the above approach. In particular, we study two additional equilibrium model systems that build upon the work presented in this manuscript: (1) hard ellipses which possess orientational, as well as positional, degrees of freedom, and (2) the Widom-Rowlinson non-additive disk model which de-mixes compositionally without undergoing large changes in density. Finally, an important advantage of standard machine learning approaches such as PCA is that equilibrium simulation data is not required, enabling the study of a known phase transition in the non-equilibrium Random Organization Model.  

\section*{Acknowledgments}
The authors thank Michael P. Howard for valuable discussions and feedback. This research was partially supported by the National Science Foundation through the Center for Dynamics and Control of Materials: an NSF MRSEC under Cooperative Agreement No. DMR-1720595 as well as the Welch Foundation (F-1696). We acknowledge the Texas Advanced Computing Center (TACC) at The University of Texas at Austin for providing HPC resources. 

\setcounter{figure}{0}
\setcounter{equation}{0}
\renewcommand\thefigure{A\arabic{figure}}
\renewcommand{\thesection}{\thepart .\arabic{section}}
\renewcommand\theequation{A\arabic{equation}}
\renewcommand{\thesubsection}{\arabic{subsection}}
\renewcommand{\thesubsubsection}{\alph{subsubsection}}


%


\end{document}